\documentclass[10pt,a4paper]{article}
\usepackage{fullpage,charter}
\usepackage{amsmath}
\usepackage{rotating}

\newcommand{\equaldef}{\stackrel{\text{\tiny def}}{=}} 

\newtheorem{definition}{Definition}

\title{Multi-Criteria Evaluation of Partitioning Schemes for Real-Time Systems}

\author{Irina Lupu\\irina.lupu@ulb.ac.be \and Pierre Courbin\\courbin@ece.fr
\and Laurent George\\lgeorge@ece.fr \and Jo\"el Goossens\\joel.goossens@ulb.ac.be}

\begin{document}
\maketitle

\begin{abstract}
In this paper we study the partitioning approach for multiprocessor real-time scheduling. This approach seems to be the easiest since, once the partitioning of the task set has been done, the problem reduces to well understood uniprocessor issues. Meanwhile, there is no optimal and polynomial solution to partition tasks on processors. In this paper we analyze partitioning algorithms from several points of view such that for a given task set and specific constraints (processor number, task set type, etc.) we should be able to identify the best heuristic and the best schedulability test. We also analyze the influence of the heuristics on the performance of the uniprocessor tests and the impact of a specific task order on the schedulability. A study on performance difference between Fixed Priority schedulers and EDF in the case of partitioning scheduling is also considered.
\end{abstract}

\section{Introduction}

\noindent Uniprocessor scheduling has been widely studied over the last few decades. However, a single processor can no longer satisfy today's computational demands, as the miniaturization of integrated circuits reaches its physical limits~\cite{limits}. Thus, a valid solution to supply sufficient resources is the use of multiprocessor platforms.

There are two main techniques for multiprocessor scheduling. Given a task set $\tau$ and a $m$-processors platform $\pi$, one can choose between:
\begin{itemize}
	\item partitioning scheduling: $\tau$ is divided into a number of disjoint subsets less than or equal to the number of processors of the platform. Each of these subsets is assigned to one processor. Uniprocessor scheduling policies are then used locally on each processor.
	\item global scheduling: at each instant $t$, the $m$ highest priority tasks are executed on the platform allowing the migration of tasks from one processor to another with the restriction that a task cannot be executed on different processors at the same time.
\end{itemize}

The two techniques are incomparable: there are task sets which are schedulable by partitioning but not by global approach and conversely~\cite{1338647}. However, today's state of the art declares that the performance of multiprocessor partitioning scheduling has not yet been overtaken by the global approach despite its considerable improvements over the last few years ~\cite{B05EDFcomparison, B05EDF_FPcomparison}. Even if, in certain scenarios, the worst case achievable utilization is the same for both techniques when using classical schedulers like EDF or Fixed Priority (FP)~\cite{ABJ01static-priorityscheduling}, in the average case, partitioning schemes seem to behave generally better than global ones. In this paper, we focus on the partitioning approach.

Partitioning a task set is equivalent to the Bin-Packing problem: how to place $n$ objects of different sizes in $m$ boxes. This problem is known to be NP-hard. The only known solution for this kind of problem is to enumerate all possible configurations and verify their correctness one by one. Suboptimal possible solutions have been proposed in the literature and are known as \textit{partitioning heuristics}. 

Previously proposed partitioning algorithms (with both FP or EDF)~\cite{Baker2008Handbook-of-Rea} are composed of a heuristic, a uniprocessor schedulability test and, very often, a task sorting criterion. Though, there does not exist an exhaustive comparative evaluation of all of them. The present state of the art compares \emph{some} partitioning algorithms based on the First Fit heuristic (a task is assigned to the first processor that can feasibly schedule it) and the decreasing utilization sorting criterion~\cite{B05EDFcomparison, B05EDF_FPcomparison}. The evaluation is usually done in terms of success ratio (the ratio of successfully scheduled task sets over all task sets considered).


Even though the success ratio is a very important evaluation criterion, other measures like the number of processors used and the available unutilized processor capacity are important since the design of real-time systems is often constrained by: the number of available processors, whether there is a risk to encounter software or hardware errors during execution, etc.~\cite{glh09}. \emph{In this research we establish some directions for deciding which heuristic, which schedulability test and which task set sorting criterion to choose according to such design constraints}.
\paragraph{This research.} Our goal is to identify the best partitioning algorithm for different scenarios; for this purpose, we combine 4 partitioning heuristics (First Fit, Best Fit, Next Fit, Worst Fit), 8 task sorting criteria (decreasing or increasing period, utilization, etc.) and 8 schedulability tests associated to FP or EDF scheduling policies. Our aim is to obtain a nearly exhaustive list of possible partitioning algorithms. Their performance will be compared with the Optimal Partitioning Algorithm (OPA in the following) according to each one of the 3 criteria: success ratio, number of processors used and available unutilized processor capacity.

Since we implemented the OPA, we also use simulation to evaluate the suboptimality of heuristics (the performance gap according to the optimal task assignment, in terms of success ratio) and the influence of this suboptimality on the performance of the tests. In this paper we also study the influence of the task sorting criteria on the schedulability of each heuristic and each test.
\paragraph{Paper Organization.} The remainder of this paper is organized as follows: in Section~2, we describe the system model and other concepts used in this paper. Section~3 presents the parameters of the study (partitioning heuristics, task sorting criteria and schedulability tests which build the partitioned scheduling algorithms). Section~4 describes the task generation methodology and Section~5 gives the results of the simulations. In Section~6, we present the conclusions and the future work in Section~7.

\section{System Model}
\noindent \normalsize In this paper we refer to the sporadic task model. A sporadic task $\tau_i$ is defined by ($C_i$, $T_i$, $D_i$) where $C_i$ is the worst-case execution time (WCET), $T_i$ is the minimum inter-arrival time (also called the period) and $D_i$ is the relative deadline (a task released at time $t$ must be
executed by its absolute deadline $t+ D_i$). A task can be instantiated an infinite number of times. An instance of a task is called a \textit{job}. 

\indent The set $\tau=\{\tau_1,\ldots ,\tau_n\}$ is composed of $n$ sporadic tasks.

\indent For FP schedulers we consider tasks in decreasing order of their priority: $\tau_1$ is the highest priority task and $\tau_i$ has higher priority than $\tau_{i+1}$.\\
We can distinguish between three kinds of sporadic task sets: implicit deadlines ($D_i = T_i, \forall i = 1,...,n$), constrained deadlines ($D_i \leq T_i, \forall i=1,...,n$) or arbitrary deadlines (no restriction on deadlines).

\indent Each task is characterized by a utilization factor $u_i \equaldef \frac{C_i}{T_i}$ and a density $\lambda_i \equaldef \frac{C_i}{\min\{D_i,T_i\}}$.

\indent Each task set is characterized by the following measures:
\begin{itemize}
 \item task set utilization $U_{\tau} \equaldef \sum_{i=1}^n u_i$ and task set density $\Lambda_{\tau} \equaldef \sum_{i=1}^n \lambda_i$;
 \item for a given $t$, the \textit{Request Bound Function} ($\operatorname{RBF}$) represents the upper bound of the work load generated by all tasks with activation times included in the interval $[0,t)$: $\operatorname{RBF}(\tau,t) \equaldef \sum_{i=1}^n \lceil \frac{t}{T_i} \rceil \cdot C_i$. An approximation for $\operatorname{RBF}$~\cite{fb05} is: $\operatorname{RBF}^{*}(\tau, t) \equaldef \sum_{i=1}^n (C_i + u_i \cdot t)$.
 \item for a given $t$, the \textit{Demand Bound Function} ($\operatorname{DBF}$) represents the upper bound of the work load generated by all tasks with activation times and absolute deadlines in the same interval $[0,t]$: $\operatorname{DBF}(\tau, t) \equaldef \sum_{i=1}^n \max \{ {0,1+ \lfloor \frac{t-D_i}{T_i} \rfloor} \} \cdot C_i$. An approximation for $\operatorname{DBF}$ ~\cite{10.1109/TC.2006.113} is: $\operatorname{DBF}^{*}(\tau, t) \equaldef \sum_{i=1}^n \mathcal{A} _i(t)$ with $\mathcal{A}_i(t) \equaldef (C_i + (t-D_i)\cdot u_i)$ if $t \geq D_i$ and 0, if not.
\end{itemize}

In the following, we consider $\pi$ a platform with $m$ identical processors: $\pi = \left\{\pi_1, \cdots, \pi_m \right\}$. By $\tau(\pi_j)$ we will denote the task subset assigned to the processor $\pi_j$.

\section{Parameters of the Study}
Our study includes \emph{all possible combinations} between a series of 4 partitioning \emph{heuristics}, 8 \emph{task sorting criteria} and 8 \emph{schedulability tests} associated with FP or EDF scheduling policies. For each combination, we evaluate the \emph{performance} according to three criteria. In this section, we introduce each of these parameters.

%
\subsection{Heuristics}
\label{Sec::Heuristics}
As mentioned previously, the Bin-Packing problem (which lies at the basis of the partitioning technique) is know to be NP-Hard. Some heuristics have been proposed in the literature in order to solve it and they are in the focus of this paper. All of them imply a sequential assignment of tasks to processors:

\begin{itemize}
	\item First Fit (FF): a task is assigned to the first processor, starting from $\pi_{1}$, which verifies the schedulability test after the assignment.
	\item Best Fit (BF): a task is assigned to the processor which verifies the schedulability test after assignment and which minimizes the remaining processor capacity.
	\item Worst Fit (WF): a task is assigned to the processor which verifies the schedulability test after assignment and which maximizes the remaining processor capacity.
	\item Next Fit (NF): a task is assigned to the first processor in the range $\{\pi_{j}, ... , \pi_m\}$ that verifies the schedulability test after the assignment ($\pi_j$ is the current processor processor). The procedure starts from $\pi_1$.
\end{itemize}

\paragraph{Impact of the Heuristic.} Heuristics can have a great impact on the solution. Consequently, it is important to know in which of the following cases the problem falls:
\begin{enumerate}
	\item we want to minimize the number of processors used;
	\item we want to provide enough time slack (i.e., leave spare capacity on processors) in order to handle system overload.
\end{enumerate}

\subsection{Task Sorting Criteria}
\label{Sec::TaskOrderCriteria}
Often, the partitioning algorithms proposed in the literature include task sorting criteria with the purpose of increasing their success ratio. In this paper we analyze the general influence of 8 such criteria \emph{increasing/decreasing}: deadline, density, period and utilization.

\subsection{Schedulability Tests}
\label{Sec::FeasibilityTests}
The schedulability tests considered in this paper are necessary and sufficient (i.e., exact) (NST) or sufficient (ST) with polynomial (P), pseudo-polynomial (PP). The tests are designed for implicit deadlines (ID), constrained deadlines (CD) or arbitrary deadlines (AD). In our study, the arbitrary deadline tests are applied to the constrained case. In the following, the name of each test represents the associated priority assignment rule followed by the initials of its authors.

\subsubsection{EDF-based schedulability tests}
For EDF two \textit{exact} and a third one only \textit{sufficient} schedulability tests are taken into account: \\

\noindent	\textbf{EDF-LL}~\cite{lw73} (NST-P-ID): $U_{\tau} \leq 1$; \\

\noindent \textbf{EDF-BHR}~\cite{bhr90} (NST-PP-AD): $\operatorname{Load}$($\tau$) $\equaldef \sup_{t \geq 0} \frac{\operatorname{DBF}(\tau,t)}{t}$ $\leq 1$. In order to reduce the number of instants $t$ to consider, we use the following expression ~\cite{glh09}: $\operatorname{Load}$($\tau$) $= \max\{U_{\tau}, \sup_{t \in [D_{\min}, P)} \frac{\operatorname{DBF}(\tau,t)}{t} \}$, where $D_{\min} \equaldef \min \{ D_1,\dots, D_n \}$ and $P \equaldef$ lcm$\{ T_1, \dots, T_n \}$ ~\cite{glh09}. As the $\operatorname{DBF}$ changes its value only at instants corresponding to absolute deadlines, $\operatorname{Load}$($\tau$) will be computed for each $t \in \bigcup_{j=1}^n \{D_j + k_j \cdot T_j, 0 \leq k_j \leq \lceil \frac{P-D_j}{T_j} \rceil -1\}$.\\

\noindent \textbf{EDF-BF}~\cite{10.1109/TC.2006.113} (ST-P-AD): $\forall \tau_i \in \tau, D_i - \operatorname{DBF}^{*}(\tau \setminus \{\tau_i\},D_i) \geq C_i$ and $1 - \sum_{\tau_j \in \tau, \tau_j \neq \tau_i}{u_j} \geq u_i$.\\

As previously mentioned, the EDF-BHR test is exact but it has a pseudo-polynomial complexity~\cite{bhr90}. A solution to this problem has been proposed~\cite{glh09}: reduce the number of analyzed instants (in the interval [$D_{\min}, P$)) by formulating the problem as a Linear Programming one; in order to solve it use the \textit{simplex} algorithm. Even though the number of instants is significantly reduced, the EDF-BF test offers a simpler (polynomial) solution to verify the schedulability of the task sets, but it is only a sufficient test.

\subsubsection{FP-Based Schedulability Tests}
FP schedulers assign fixed priorities to tasks before the execution of the system and the jobs inherit the priority of the task that generated them. During our simulations we consider DM (for AD et CD tests) and RM (for ID tests) schedulers~\cite{lw73,Dd90deadlinemonotonic}.\newline

\noindent \textbf{DM-ABRTW}~\cite{auds93} (NST-PP-CD): DM-test based on the response-time analysis: $\forall \tau_i \in \tau$, $r_i \leq D_i$, where $r_i$ is $\tau_i$'s \textit{worst case response time}. For constrained deadlines, $r_i$ is determined for the first activation of the task in the synchronous scenario as a solution of the equation: $r_i = C_i + \sum_{j = 1}^{i-1} \lceil \frac{r_i}{T_j}\rceil \cdot C_j$ computed as follows: $W_0 = C_i$ and $W_{k+1} = C_i + \sum_{j = 1}^{i-1} \lceil \frac{W_k}{T_j}\rceil \cdot C_j$ until $W_{k+1} = W_k$ or $W_k > D_i$.\newline 

\noindent \textbf{DM-FBB}~\cite{1153942} (ST-P-AD): $\forall \tau_i \in \tau, D_i - \operatorname{RBF}^{*}(\tau - \tau_i,D_i) \geq C_i$ and $1 - \sum_{\tau_j \in \tau, \tau_j \neq \tau_i}{u_j}  \geq u_i$. \newline

\noindent \textbf{RM-LL}~\cite{lw73,goosdevLL} (ST-P-ID): $U_{\tau} \leq n(\sqrt[n]{2} - 1)$.\newline

\noindent \textbf{RM-BBB}~\cite{10.1109/TC.2003.1214341} (ST-P-ID): $\prod_{i=1}^n{(u_i + 1)} \leq 2$. \newline

\noindent \textbf{RM-LMM}~\cite{lmm98} (ST-P-ID): If $1 \leq r_{\pi} \equaldef \frac{T_{\operatorname{max}}(\tau)}{T_{\operatorname{min}}(\tau)} < 2$\\ (where $T_{\operatorname{max}}(\tau) \equaldef \max{\left\{T_1, \ldots, T_n \right\}}$ and $T_{\operatorname{min}}(\tau) \equaldef \min{\left\{T_1, \ldots, T_n \right\}}$), then $\tau$ is RM schedulable on processor $\pi$ if $U_{\tau} \leq B(r_{\pi}, n)$ with $B(r_{\pi}, n) \equaldef n \cdot (\sqrt[n]{r_{\pi}} -1) + \frac{2}{r_{\pi}} -1$. For the case $r_{\pi} \notin [1,2)$ a scaling algorithm is presented in~\cite{lmm98} such that, if $\tau'$ is the task set obtained after scaling: $\tau$ is schedulable if and only if $\tau'$ is schedulable.\newline


DM-ABRTW is an exact test, but it has a pseudo-polynomial complexity. DM-FBB is a uniprocessor schedulability test which has been proposed especially for partitioned scheduling ~\cite{1153942}. Simulations show that when DM-ABRTW and DM-FBB are associated with the FF heuristic, their behavior is incomparable: for constrained deadline task sets DM-ABRTW behaves better, but for the case of arbitrary deadlines, DM-FBB has a higher success rate than DM-ABRTW. 

RM-LL is the popular Liu \& Layland FP test. For a sufficiently large number of tasks in the set, this test has an utilization bound of $\ln 2$ ($\cong 0.69$) ~\cite{lw73}. However, RM can generally schedule task sets with a utilization of 88\% (this shows the pessimism of the RM-LL test). RM-BBB test is also polynomial, but it establishes a higher task set utilization bound than RM-LL. Simulations show~\cite{10.1109/TC.2003.1214341} that RM-BBB schedules $\sqrt{2}$ more task sets than RM-LL as the number of tasks grows. The last FP schedulability test, RM-LMM has already been applied in multiprocessor scheduling and compared with RM-LL when associated with the FF heuristic ~\cite{lmm98}. Simulations show that RM-LMM's average processor utilization is comparable to that of an exact test: 96\%. In the same scenario, the RM-LL average processor utilization is 75\%.\\

For BF and WF partitioning heuristics, the schedulability tests determine the expression used to compute the unutilized processor capacity when tasks are assigned to processors: $1 - \Lambda_{\tau}$ for EDF-LL, EDF-BF, DM-ABRTW, DM-FBB, RM-LL, RM-BBB and RM-LMM; and $1 - \operatorname{Load}(\tau)$ for EDF-BHR, since this is the only test based on $\operatorname{Load(\tau)}$.

\subsection{Performance Criteria}
\label{Sec::PerformanceCriteria}
Each combination of the previously mentioned parameters is evaluated according to several performance criteria:

\begin{itemize}
\item \textit{Success Ratio}: with this criterion we determine which combination schedules the largest number of task sets. It is defined as follows:

\vspace{0.1cm}
\hspace*{0.5cm}\noindent $\frac{\operatorname{number\ of\ task\ sets\ successfully\ scheduled}}{\operatorname{total\ number\ of\ task\ sets}}$

\item \textit{Number of processors used} defined as number of processors where at least one task is assigned if $m$ is sufficient.

\item \textit{Average value of processor spare capacity}: the spare capacity on processors is computed with the expression $1-\operatorname{Load}(\tau(\pi_j))$ for EDF-BHR and $1-\Lambda_{\tau(\pi_j)}$ for the other schedulability tests.
\end{itemize} 

\subsection{Optimal Partitioning Algorithm}
\label{Sec::Optimal}
The OPA is designed as follows: enumerate all possible task assignments on processors and test them one by one.
In terms of schedulability tests, for EDF, we associate this optimal assignment with the exact test of EDF-BHR ($\operatorname{OPA[EDF]}$). For FP, we use the exact test of DM-ABRTW ($\operatorname{OPA[FP]}$).\\

\section{Task Generation Methodology}
The task generation methodology used in this paper is based on the one presented in~\cite{B05EDFcomparison}. However, in our case, task generation is adapted to each type of deadline considered. In the following, $k_i \in \{D_i, T_i\}$ and $\rho_i \in \{u_i, \lambda_i\}$:
\begin{itemize}
	\item $k_i$ is uniformly chosen from [1,100];
	\item $\rho_i$ (truncated between 0.001 and 0.999) has the following distributions:
	\begin{enumerate}
		\item Uniform distribution between $\frac{1}{k_i}$ and 1;
		\item Bimodal distribution: heavy tasks have a uniform distribution between 0.5 and 1, light tasks have a uniform distribution between $\frac{1}{k_i}$ and 0.5; the probability of a task being heavy is $\frac{1}{3}$;
		\item Exponential distribution of mean 0.25;
		\item Exponential distribution of mean 0.50.
	\end{enumerate}
\end{itemize}
 \normalsize
For implicit deadline task sets, $(k_i,\rho_i) = (T_i, u_i)$ and for constrained deadlines $(k_i,\rho_i) = (D_i, \lambda_i)$. In this last case, the period $T_i$ is uniformly chosen from $[D_i, 100]$. \newline
\indent We consider 4-processor identical platforms. \newline
\indent Task systems are generated so that those obviously unfeasible ($U_{\tau} > m$) or trivially schedulable ($m=n$ and $\forall i \in [1,n]$, $u_i \leq 1$, meaning the capacity of one processor of the identical platform) are not considered during simulations: \newline
\indent Step~1: Initially, we generate a system which contains $m+1$ tasks and test it. \newline
\indent Step~2: We add tasks to the system and repeat the tests until the density of the system exceeds $m$ (the capacity of the identical platform).

For our simulations, we generated ¨$10^6$ task sets uniformly chosen from the distributions mentioned above with implicit and constrained deadlines.  

\section{Results}
This section presents a comparative study of several partitioning scheduling algorithms with regard to the criteria introduced in Section \ref{Sec::PerformanceCriteria}.\\

\noindent The evaluation is structured as follows:

\begin{enumerate}
\item We study the suboptimality of FP over EDF in terms of success ratio in a multiprocessor environment.
\item In the same way, we evaluate the suboptimality of each heuristic compared with the optimal placement.
\item We determine the success ratio of each test when associated with partitioning heuristics.
\item For each given test, we determine the sorting criterion that maximizes its schedulability when associated with partitioning heuristics.
\item We compare the success ratio, number of processors used and available spare capacity of all heuristics (all tests and task set sorting criteria included).
\item Based on the best heuristic determined previously, we find the best couple heuristic-task sorting criterion.
\end{enumerate}


Firstly, we have to define the concept of \textit{suboptimality degree}. The suboptimality degree will be computed as a function of two parameters, $p_1$ and $p_2$. It is defined as follows:

\begin{definition}[Suboptimality Degree]
The degree by which the success ratio of $p_1$ is overpassed by the one of $p_2$ ($p_2 \geq p_1$ and $p_2 > 0$):

\vspace{0.2cm}
$\operatorname{sd(p_1,p_2) = \frac{SuccessRatio\ p_2\ -\ SuccessRatio\ p_1}{SuccessRatio\ p_2}\cdot100}$.\upshape \\

Smaller the value of $\operatorname{sd}(p_1,p_2)$, better the performance of $p_1$ according to the one of $p_2$.

\end{definition}

\subsection{Suboptimality of FP over EDF}
\label{Sec::SubotimalityFPEDF}
\begin{figure}[!t]
	\centering
   \includegraphics[scale=0.36]{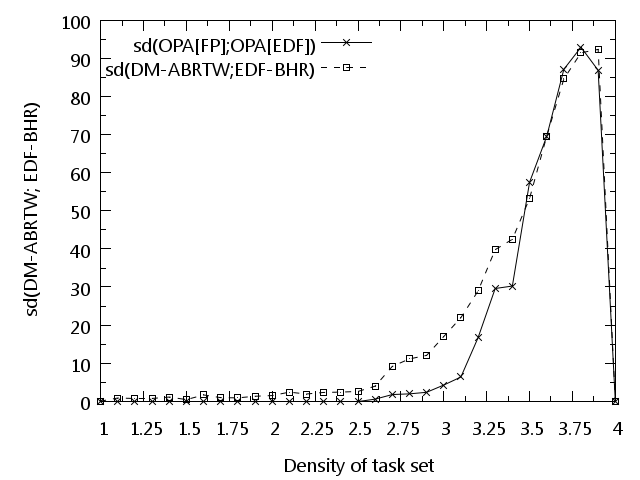}
   \caption{Heuristic - FP/EDF Suboptimality}
   \label{Graph::Results::5.3InfluenceOfHeuristics}
\end{figure}

The degree of suboptimality of FP schedulers according to EDF has been previously analyzed in the uniprocessor case \cite{drsa08}. Our study determines this degree for the multiprocessor scenario (through simulations) in relation with the total density of the task set. Figure \ref{Graph::Results::5.3InfluenceOfHeuristics} shows the simulation results as follows:

\hspace*{0.2cm} (1) ~$\operatorname{sd(OPA[FP],OPA[EDF])}$: the suboptimality degree for the case of an optimal task assignment;

\hspace*{0.2cm} (2) ~$\operatorname{sd(DM-ABRTW,EDF-BHR)}$: the suboptimality degree for the case where the same exact tests are combined with all 4 heuristics.

\vspace{0.1cm}
The maximum gain of EDF over FP in the case of the optimal task assignment is 93\% for high total density task sets.

When the exact tests are associated with the 4 heuristics the suboptimality degree of FP over EDF slightly increases. Though, the two curves have generally the same shape which means that the heuristics do not influence importantly the suboptimality degree of the schedulability tests.


\subsection{Suboptimality of Heuristics}
\label{SubSection::SubOptimalityOfHeuristics}

\begin{figure}[!t]
	\centering
   \includegraphics[scale=0.36]{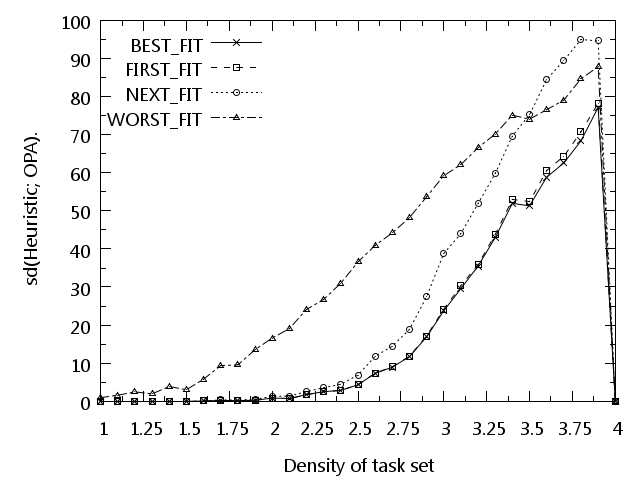}
   \caption{Suboptimality of heuristics}
   \label{Graph::Results::5.2}
\end{figure}

By definition, a heuristic is a suboptimal solution. In this section, we present the suboptimality degree of each heuristic according to the optimal task assignment. The associated schedulability test is EDF-BHR and the simulations results include all sorting criteria. Figure \ref{Graph::Results::5.2} shows the results computed as follows:

\vspace{0.1cm}
\hspace*{0.7cm} $\operatorname{sd(Heuristic, OPA)}$.

First of all, we should mention that, in terms of complexity, the four heuristics can be listed in decreasing order as follows: BF and WF (equal complexities), FF and finally, NF.

Figure \ref{Graph::Results::5.2} shows that for task sets with the total density bounded by half the capacity of the platform, BF, FF and NF perform the same. As NF is the least complex, it is more convenient to choose it in this case. For the scenario where the total density exceeds half of the platform capacity, BF is the best choice. Taking into account the very slight difference between FF and BF and the fact that FF has a lower complexity, FF should be also considered.

\subsection{Choosing a schedulability test}

In this section, we analyze the success ratio of schedulability tests for all possible combinations with the 4 heuristics and the 8 sorting criteria. The analysis is divided in three subsections: firstly, EDF tests, secondly, FP tests and finally, a comparison between the best performers of EDF and FP.

\subsubsection{EDF}

For \emph{implicit deadlines}, all EDF tests reduce to EDF-LL. 

\begin{figure}[!t]
	\centering
   \includegraphics[scale=0.36]{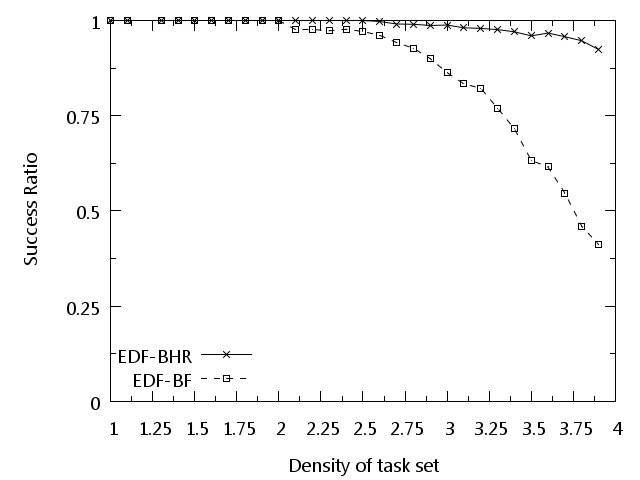}
   \caption{EDF - Constrained Deadlines}
   \label{Graph::Results::5.3.1EDFConstrained}
\end{figure}

\vspace{0.1cm}
\noindent For the case of \emph{constrained deadlines} and total task set density inferior to half of the platform capacity, the two tests have the same performance as seen in Figure \ref{Graph::Results::5.3.1EDFConstrained}.

\noindent So, EDF-BF is the best option in this case because of its polynomial complexity. For the case with the total density exceeding half of the platform capacity, EDF-BHR is a better choice despite its pseudo-polynomial complexity.


\subsubsection{Fixed Priority}

\begin{figure}[!t]
	\centering
   \includegraphics[scale=0.36]{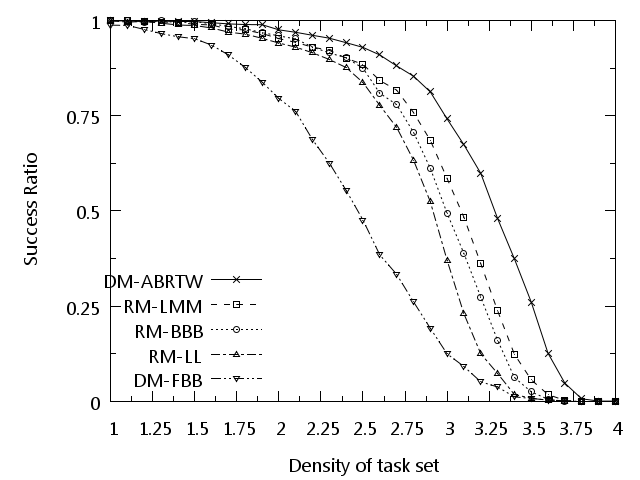}
   \caption{FP - Implicit Deadlines}
   \label{Graph::Results::5.3FPImplicit}
\end{figure}

For \emph{implicit deadlines}, all the FP tests were taken into account during simulations. As DM-ABRTW is an exact test, it has the best performance even when associated with heuristics. For the sufficient tests, we can identify a decreasing order of their performance: RM-LMM is the best sufficient test followed by RM-BBB (which outperforms RM-LL as in the uniprocessor case); and, RM-LL performs better than DM-FBB, as is shown in Figure \ref{Graph::Results::5.3FPImplicit}.

\begin{figure}[!t]
	\centering
   \includegraphics[scale=0.36]{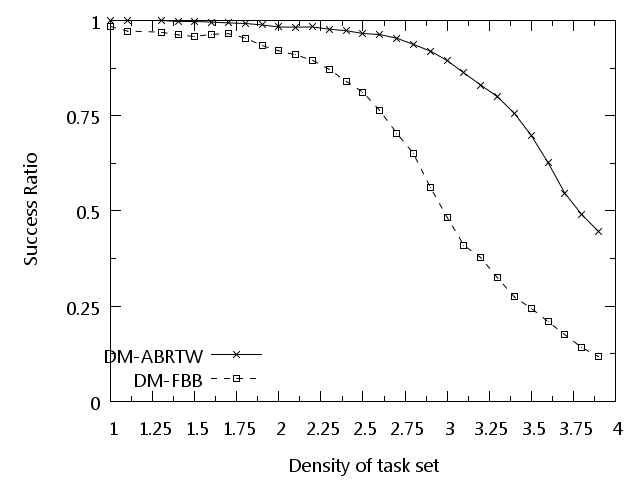}
   \caption{FP - Constrained deadlines}
   \label{Graph::Results::5.3FPConstrained}
\end{figure}

\vspace{0.1cm}
For the \emph{constrained deadlines} case, DM-ABRTW is an exact schedulability test which, naturally, performs better than the only sufficient DM-FBB (Figure \ref{Graph::Results::5.3FPConstrained}). 

\subsection{Choosing a Task Sorting Criterion}

\begin{figure}[!t]
	\centering
   \includegraphics[scale=0.36]{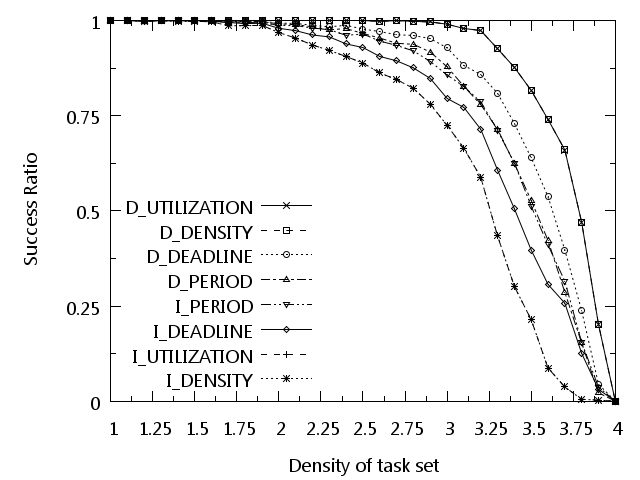}
   \caption{EDF-based Sorting Criteria}
   \label{Graph::Results::5.4ChooseEDFSortingCriteria}
\end{figure}

This section deals with the impact of a task sorting criterion on the success ratio of schedulability tests. In the corresponding graphs, \emph{D} means \emph{decreasing} and \emph{I} means \emph{increasing}. Figure~\ref{Graph::Results::5.4ChooseEDFSortingCriteria} shows for EDF-BHR test the sorting criteria that maximize its success ratio: Decreasing Utilization and Decreasing Density.

The simulation results showed that all tests (based either on EDF or FP and excluding DM-FBB) have the same sorting criteria which maximizes their success ratio as EDF-BHR. For this reason and due to space constraints, we did not add the corresponding graphs in this paper.

\begin{figure}[!t]
	\centering
   \includegraphics[scale=0.36]{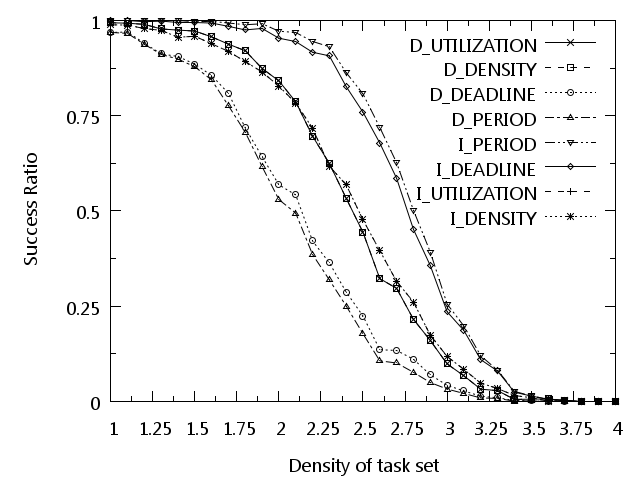}
   \caption{FBB Sorting Criteria}
   \label{Graph::Results::5.4ChooseFBBSortingCriteria}
\end{figure}

DM-FBB test performs the best when associated with Period or Deadline Increasing order as seen in Figure \ref{Graph::Results::5.4ChooseFBBSortingCriteria} with a better performance when associated with Period Increasing order.

\subsection{Choosing an Heuristic}
\label{SubSection::Heuristic}

In this section we evaluate the performance of the heuristics according to certain evaluation criteria. In this analysis each heuristic is combined with all the schedulability tests and the task sorting criteria.

\begin{figure}[!t]
	\centering
   \includegraphics[scale=0.36]{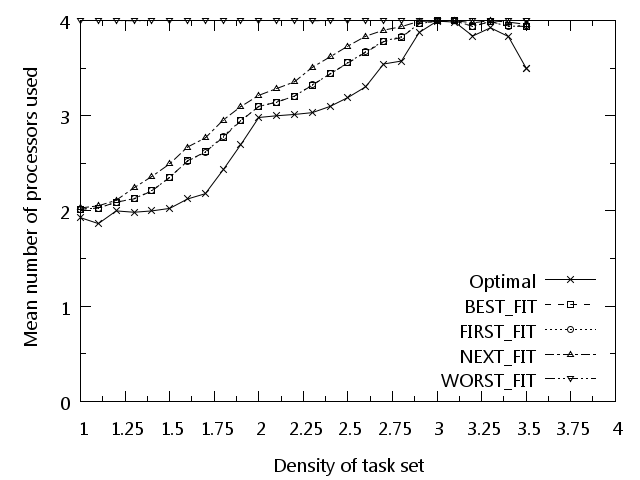}
   \caption{Minimum number of processors}
   \label{Graph::Results::5.5ProcessorsUtilized}
\end{figure}

\vspace{-0.2cm}
\paragraph{Minimum Number of Used Processors.}
As seen in Figure \ref{Graph::Results::5.5ProcessorsUtilized}, the heuristic that uses the smallest number of processors is FF and the one that uses the largest is WF. 

\begin{figure}[!t]
	\centering
   \includegraphics[scale=0.36]{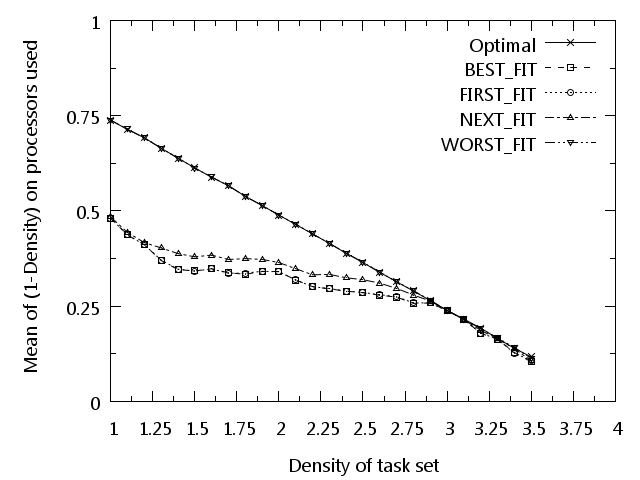}
   \caption{Unutilized capacity - $1-\Lambda_{\tau}$}
   \label{Graph::Results::5.5CapacityOfProcessorsU}
\end{figure}

\vspace{-0.2cm}
\paragraph{Available Spare Capacity on Processors.} As WF utilizes the maximum number of processors, the available spare capacity is also maximized. Figure \ref{Graph::Results::5.5CapacityOfProcessorsU} shows that WF behaves as an optimal placement according to the $1-\Lambda_{\tau}$ criterion.

\begin{figure}[t]
	\centering
	\includegraphics[scale=0.36]{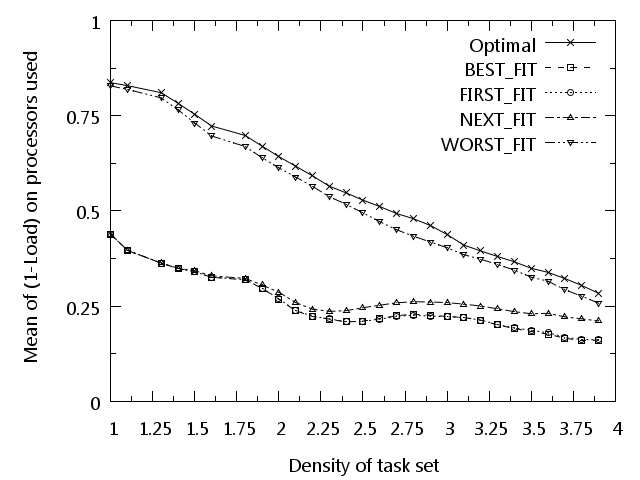}
  \caption{Unutilized capacity - $1-$\textit{Load}($\tau$)}
  \label{Graph::Results::5.5CapacityOfProcessorsLoad}
\end{figure}

\noindent Also for the $1-\operatorname{Load}({\tau})$ criterion, WF has the closest behavior to the optimal task assignment, as shown in Figure~\ref{Graph::Results::5.5CapacityOfProcessorsLoad}.

In Section \ref{Sec::Heuristics}, we presented two cases for multiprocessor scheduling problem. According to the simulation results presented above, we can conclude:
\begin{itemize}
	\item if we want to minimize the number of used processors, the best heuristic is BF which slightly outperforms FF.
	\item if we want to ensure an execution time slack (for the case where there is a risk to encounter software or hardware errors), the most suitable heuristic is WF with a behavior close to the one of an optimal assignment.
\end{itemize}

\begin{figure}[t]
	\centering   
   \includegraphics[scale=0.36]{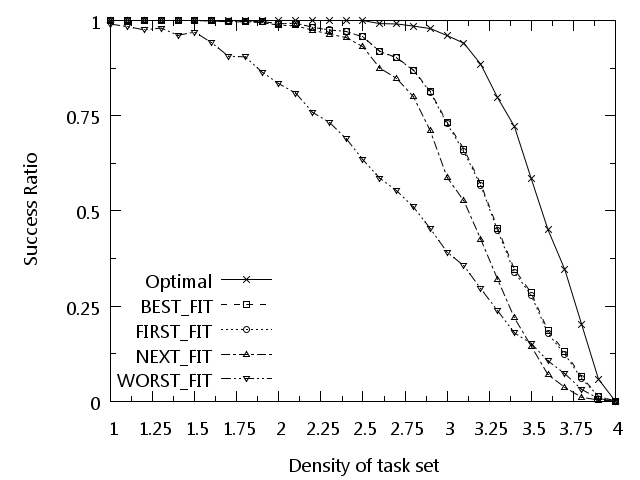}
   \caption{Heuristic - Success Ratio}
   \label{Graph::Results::5.5SuccessRatio}
\end{figure}

\paragraph{Success Ratio.} In Figure \ref{Graph::Results::5.5SuccessRatio} we can observe that the success ratio of partitioning heuristics (when combined with all the schedulability tests and all the sorting criteria) follows the same performance order as in Figure \ref{Graph::Results::5.2}: BF, FF, NF, WF. Taking into account the complexity of the heuristics and the density of the task set, we can choose: NF, if task set requires no more than 50\% of the platform capacity for execution (due to its low complexity) or, if the task set requires more than this 50\% bound, BF should be used for task assignment on processors.

\subsection{Choosing a task sorting criteria for the best heuristic}

According to Section \ref{SubSection::Heuristic}, the best heuristic are BF and FF with the same success ratio. Due to its low complexity, FF is usually considered when designing partitioning algorithms. It is generally agreed that the best association partitioning heuristic --- sorting criterion is FFDU (First Fit Decreasing Utilization).

\begin{figure}[t]
	\centering
   \includegraphics[scale=0.36]{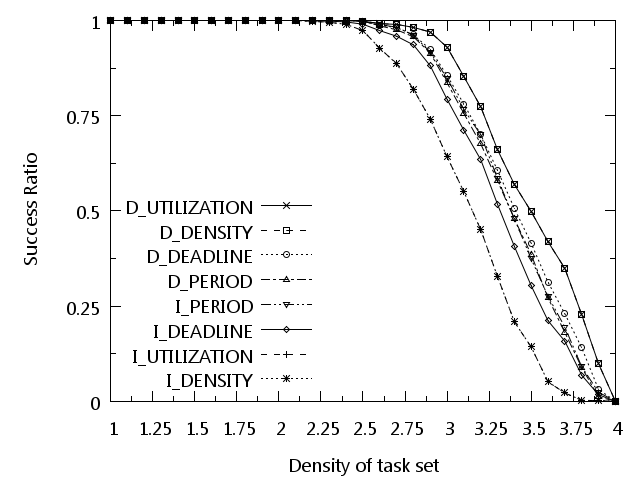}
   \caption{Task sorting criteria for FF}
   \label{Graph::Results::5.6ChooseTaskOrderForBestHeuristic}
\end{figure}

Figure \ref{Graph::Results::5.6ChooseTaskOrderForBestHeuristic} shows that for task sets with the total density inferior to 75\% of the platform capacity, all sorting criteria give the same performance. However, for task sets with total density higher than 75\% of the platform capacity, Decreasing Utilization (for implicit deadlines) and Decreasing Density (for constrained deadlines) exhibit the best behavior. 

\section{Conclusion}\label{Section_Conclusion}

The main conclusion of our study is that given a task set and a processor platform, the schedulability test, the heuristic and the sorting criteria that compose the partitioning algorithm, have to be chosen according to:
\begin{itemize}
	\item the total task density;
	\item the concern to:
\begin{itemize}
	\item minimize the number of processors;
	\item ensure an execution time slack.
\end{itemize}
\end{itemize}

For the choice of schedulability tests, we tend to select exact ones due to their better performance. Furthermore, given that EDF dominates any FP scheduler, we are likely to consider EDF. However, when choosing a schedulability test, it is important to observe the task set total density, as, for values bounded by half of the platform capacity, exact tests perform the same as sufficient ones which is an advantage given the lower implementation complexity.

The task set total density is also important when choosing an heuristic: in the case of minimizing the number of processors, for task sets with density bounded by half of the platform capacity, NF is the best choice, but for task sets with total density higher than 50\% of the platform capacity, BF has to be considered. Furthermore, we want to ensure an execution slack, WF has a behavior close to the one of optimal task assignment.

From the simulations, we have deducted that the best task sorting criteria is Decreasing Density for all schedulability tests, even if the priority assignment is based on the deadline (EDF, DM) or the period (RM). There is one exception, DM-FBB which has the best performance when associated to Increasing Period sorting criteria.

According to the assertions above, one can guide its choices regarding the test, the heuristic and the task sorting criteria for scheduling a real-time task set on a multiprocessor platform.

\section{Future Work}\label{Section_FutureWork}
In the future, we firstly want to extend this study to heterogeneous platforms. Secondly, using the same approach as for this analysis, we want to evaluate the performance of a new class of multiprocessor algorithms called \emph{semi-partitioning}. As this technique combines the 2 previously existing ones (global and partitioning), it would be interesting to widen the analysis with new evaluation criteria such as the number of context switches (migrations and preemptions).
\bibliographystyle{acm}
\bibliography{Bibliographie/Bibliographie}

\begin{thebibliography}{10}\setlength{\itemsep}{-1ex}\small

\bibitem{ABJ01static-priorityscheduling}
B.~Andersson, S.~Baruah, and J.~Jonsson.
\newblock Static-priority scheduling on multiprocessors.
\newblock In {\em 22nd IEEE Real-Time Systems Symposium}, pages 193--202, 2001.

\bibitem{auds93}
N.~Audsley, A.~Burns, M.~Richardson, K.~Tindell, and J.~Wellings.
\newblock Applying new scheduling theory to static priority pre-emptive
  scheduling.
\newblock {\em Software Engineering Journal}, 8:284--292, 1993.

\bibitem{Dd90deadlinemonotonic}
N.~C. Audsley, A.~Burns, M.~F. Richardson, and A.~J. Wellings.
\newblock Real-time scheduling: the deadline-monotonic approach.
\newblock In {\em IEEE Workshop on Real-Time Operating Systems and Software},
  pages 133--137, 1991.

\bibitem{B05EDF_FPcomparison}
T.~P. Baker.
\newblock Comparison of empirical success rates of global vs. partitioned
  fixed-priority and {EDF} scheduling for hard real time.
\newblock Technical Report TR-050601, The Florida State University, 2005.

\bibitem{B05EDFcomparison}
T.~P. Baker.
\newblock A comparison of global and partitioned {EDF} schedulability tests for
  multiprocessors.
\newblock In {\em International Conf. on Real-Time and Network Systems}, pages
  119--127, 2006.

\bibitem{Baker2008Handbook-of-Rea}
T.~P. Baker and S.~Baruah.
\newblock {\em Handbook of Real-Time and Embedded Systems}, chapter
  Schedulability Analysis of Multiprocessor Sporadic Task Systems, pages
  3.1--3.15.
\newblock Chapman, 2008.

\bibitem{1338647}
S.~Baruah.
\newblock Techniques for multiprocessor global schedulability analysis.
\newblock In {\em 28th IEEE International Real-Time Systems Symposium}, pages
  119--128, 2007.

\bibitem{10.1109/TC.2006.113}
S.~Baruah and N.~Fisher.
\newblock The partitioned multiprocessor scheduling of deadline-constrained
  sporadic task systems.
\newblock {\em IEEE Trans. on Computers}, 55(7):918--923, 2006.

\bibitem{bhr90}
S.~Baruah, R.~Howell, and L.~Rosier.
\newblock Algorithms and complexity concerning the preemptive scheduling of
  periodic real-time tasks on one processor.
\newblock {\em Real-Time Systems}, 2:301--324, 1990.

\bibitem{10.1109/TC.2003.1214341}
E.~Bini, G.~C. Buttazzo, and G.~M. Buttazzo.
\newblock Rate monotonic analysis: The hyperbolic bound.
\newblock {\em IEEE Transactions on Computers}, 52(7):933--942, 2003.

\bibitem{drsa08}
R.~Davis and A.~Burns.
\newblock Exact quantification of the sub-optimality of uniprocessor
  fixed-priority pre-emptive scheduling.
\newblock {\em Real-Time Systems}, (3):211--258, 2009.

\bibitem{goosdevLL}
R.~Devillers and J.~Goossens.
\newblock {Liu} and {Layland}'s schedulability test revisited.
\newblock {\em Information Processing Letters}, 73(5--6):157--161, 2000.

\bibitem{1153942}
N.~Fisher, S.~Baruah, and T.~Baker.
\newblock The partitioned scheduling of sporadic tasks according to
  static-priorities.
\newblock In {\em Proceedings of the 18th Euromicro Conference on Real-Time
  Systems}, pages 118--127, Dresden, Germany, 2006.

\bibitem{fb05}
N.~Fisher and S.~K. Baruah.
\newblock A fully polynomial-time approximation scheme for feasibility analysis
  in static-priority systems with arbitrary relative deadlines.
\newblock In {\em Proceedings of the 17th Euromicro Conference on Real-Time
  Systems}, pages 117--126, 2005.

\bibitem{glh09}
L.~George and J.~Hermmant.
\newblock {A norm approach for Partitioned EDF Scheduling of Sporadic Task
  Systems}.
\newblock {\em Proceedings of the 21st Euromicro Conference on Real-Time
  Systems}, Dublin, Ireland, July 2009.

\bibitem{lmm98}
S.~Lauzac, R.~G. Melhem, and D.~Moss{\'e}.
\newblock An efficient {RMS} admission control and its application to
  multiprocessor scheduling.
\newblock In {\em 12th International Parallel Processing Symposium}, pages
  511--518, 1998.

\bibitem{lw73}
L.~C. Liu and W.~Layland.
\newblock Scheduling algorithms for multi-programming in a hard real time
  environment.
\newblock {\em Journal of ACM}, 20(1):46--61, January 1973.

\bibitem{limits}
V.~Zhirnov, R.~Cavin, J.~Hutchby, and G.~Bourianoff.
\newblock Limits to binary logic switch scaling--a {G}edanken model.
\newblock In {\em IEEE}, volume~9, pages 1934--1939, 2003.

\end{thebibliography}
\end{document}